\documentclass[dvips]{article}
\usepackage{icrctc07}

\title{Positrons from dark matter annihilation in the galactic halo: uncertainties}
\shorttitle{Positrons from dark matter annihilation in the galactic halo: uncertainties}

\authors{N. Fornengo$^{1}$, T. Delahaye$^{2}$, R. Lineros$^{1}$, F. Donato$^{1}$, P. Salati$^{2}$}
\shortauthors{Fornengo et al.}
\afiliations{$^1$Universit\`a degli Studi di Torino and Istituto Nazionale di Fisica Nucleare, Sezione di Torino, via P. Giuria 1, I--10125 Torino, Italy\\ 
$^2$Laboratoire d'Annecy-le-Vieux de Physique Th\'eorique LAPTH, CNRS-SPM and Universit\'e de Savoie 9, Chemin de Bellevue, B.P.110 74941 Annecy-le-Vieux, France}
\email{fornengo@to.infn.it,delahaye@lapp.in2p3.fr,lineros@to.infn.it,donato@to.infn.it,salati@lapp.in2p3.fr}

\abstract{Indirect detection signals from dark matter annihilation are studied in the positron channel. We discuss in detail the positron propagation inside the galactic medium: we present novel solutions of the diffusion and propagation equations and we focus on the determination of the astrophysical uncertainties which affect the positron dark matter signal. We show that, especially in the low energy tail of the positron spectra at Earth, the uncertainty is sizeable and we quantify the effect. Comparison of our predictions with current available and foreseen experimental data are derived.}

\begin{document}
\maketitle

\section{Introduction}

The quest for the identification of dark matter (DM), together with the comprehension of the nature of dark energy, is one of the most important and potentially far--reaching problems in the understanding of the physical world, and represents a matter of investigation where cosmology, astrophysics and particle physics converge together. It is therefore of utmost importance to address the problem of the identification of DM from many different points of view, which bring together research performed, from one side, in particle accelerators and on the other side in underground laboratories, in neutrino telescopes, in large--area surface detectors as well as in space. In this paper we attack the problem from the point of view of the indirect detection of DM through the search of its products of annihilation inside the galactic halo. We will specify our discussion on the positron signal, for which we intend to quantify the astrophysical uncertainties and derive the ensuing consequences for upcoming experiments. A more extended and detailed discussion of our analyses will be presented elsewhere \cite{positrons}.

\section{The diffusion equation and its solution}

\begin{figure}[t]
\begin{center}
\noindent
\includegraphics [width=0.44\textwidth]{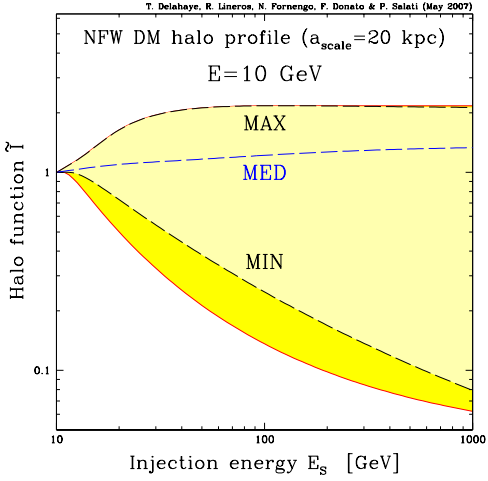}
\end{center}
\vskip -0.75cm
\caption{
The halo function $\tilde{\cal{I}}$ is plotted as a function of the positron injection energy $E_{S}$,
for an observed energy of 10 GeV. The galactic halo profile is NFW \cite{nfw} with a scale radius of
20 kpc. The curve labeled as MED corresponds to the choice of propagation parameters which best--fit
the B/C ratio \cite{parfit}. The MAX and MIN configurations correspond to the cases which were identified
to produce the maximal and minimal DM antiproton fluxes \cite{antiprotons}, while the larger area denotes
the full variation of the propagation parameters yet compatible with the B/C analysis \cite{parfit}.
}
\label{fig1}
\end{figure}

The propagation of positrons in the galactic medium is governed by the transport equation:
\begin{displaymath}
-K(E) \nabla^{2}\Psi(E) - \frac{\partial}{\partial E}\big\{b(E) \Psi(E)\big\} = Q(\mathbf{x},E) ,
\end{displaymath}
where $\Psi(E)$ is the positron number density per unit energy and $Q(\mathbf{x},E)$ is the positron source term.
The transport through the magnetic turbulences is described by the space independent diffusion coefficient
$K(\mathbf{x},E) \! = \! K_0 \, \epsilon^\delta$ where $\epsilon \! = \! {E}/{E_0}$ and $E_{0} \! = \! 1$ GeV.
Positrons lose energy through synchrotron radiation and inverse Compton scattering on the cosmic microwave background radiation
and on the galactic starlight at a rate $b(E) \! = \! {E_0} \, {\epsilon^2} / {\tau_E}$ where $\tau_E \! = \! 10^{16}$ s.
The source term for positrons produced through DM annihiliation is then:
\begin{displaymath}
Q(\mathbf{x},E) =  \alpha \, \frac{\langle\sigma v\rangle}{m^2_\chi} \ f(E) \ \rho^{2}(\mathbf{x}) =
\kappa \ f(E) \ \frac{\rho^2(\mathbf{x})}{\rho_{\odot}^2} ,
\end{displaymath}
where $m_{\chi}$ is the mass of the DM particle and $\langle\sigma v\rangle$ is the low--energy velocity--averaged DM annihilation cross--section.
The positron spectrum $f(E)$ produced in the annihilation process depends on the specific final states of the annihilation cross section.
The DM density profile of the galactic halo is denoted by $\rho(\mathbf{x})$ and its solar value by $\rho_{\odot}$ whereas
$\alpha \! = \! 1/2$ (Majorana) or $1/4$ (Dirac).

The transport of positrons through the Galaxy takes place into a diffusion zone (DZ), which we model as a cylinder with medium height $L_z$ and
radius $R_{g} \! = \! 20$ kpc. Until now, the diffusion equation has been solved using the
method of Green functions and we have improved that technique with the help of Bessel functions
which allow to take naturally into account the galactic radial boundaries (for details, see \cite{positrons}).
The positron density may be expressed in terms of a halo function $\tilde{\cal{I}}$ as:
\begin{displaymath}
\Psi(E) =  \kappa \; \frac{\tau_E }{E_0 \epsilon^2} \int_E^{\infty}  dE_S
\ f(E_S) \ \tilde{\cal{I}}(\lambda_D) .
\end{displaymath}
The typical positron diffusion scale $\lambda_D$ depends on the injection energy $E_{S}$
and on the observed energy $E$ as shown in \cite{lavalle_06}.
That halo function may be written as:
\begin{displaymath}
\tilde{\cal{I}}(\lambda_D) = \sum_g C_g \ \chi_g({\mathbf{x}}_{\odot})  \exp{\left(-\frac{g^2}{4} \lambda_D^2\right)} \ ,
\end{displaymath}
where
\begin{displaymath}
C_g = \frac{1}{\pi R_{g}^2 L_z} \,
\int_{\textnormal{DZ}} d^3{\mathbf{x}} \, \frac{\rho^2(\mathbf{x})}{\rho_{\odot}^2} \, \chi^{\dagger}_g(\mathbf{x}) ,
\end{displaymath}
and $\chi_g$ are eigenfunctions of the Helmholtz equation \cite{positrons}.
The halo function $\tilde{\cal{I}}$ is featured in Fig.\ref{fig1} as a function of the injection energy
$E_{S}$ for a fixed value of the observed positron energy $E$ and assuming a NFW density profile \cite{nfw}.
In order to quantify the uncertainty related to the properties of the diffusion region, we have varied
the astrophysical parameters which govern the diffusion process inside the range allowed by the B/C measurements \cite{parfit}.
Fig.\ref{fig1} shows that, for large injection energies as compared to the observed energy,
the uncertainty on the halo function is sizeable, reaching a factor of two upwards and of
20 downwards at $E_{S} \! = \! 1$ TeV. Other DM profiles will be discussed in \cite{positrons}.
%

\section{The positron signal}

\begin{figure}[t]
\begin{center}
\noindent
\includegraphics [width=0.44\textwidth]{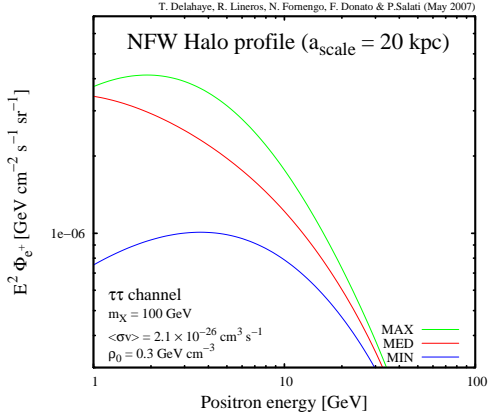}
\end{center}
\vskip -0.75cm
\caption{Positron fluxes for a 100 GeV DM particle which annihilates into a $\bar{\tau}\tau$ final state, as a function of the positron energy. The galactic halo profile is NFW \cite{nfw} with a scale radius of 20 kpc. The upper, median and lower curves show the effect due to galatic propagation: the median curve refers to the best--fit set of propagation parameters \cite{parfit}, while the upper and lower curves delimit the uncertainty band on the theoretical determination of the flux.}\label{fig2}
\end{figure}

\begin{figure}[t]
\begin{center}
\noindent
\includegraphics [width=0.44\textwidth]{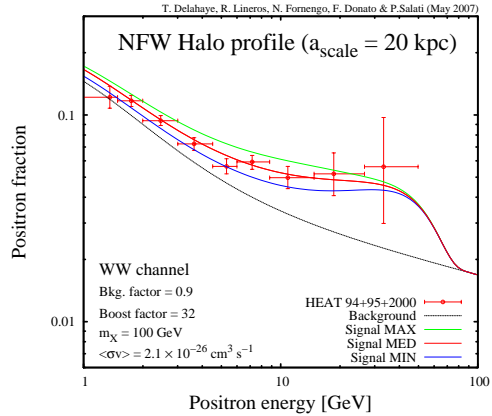}
\end{center}
\vskip -0.75cm
\caption{Positron fraction $e^{+}/(e^{+}+e^{-})$ as a function of the positron energy. The lower solid curve refers to the background estimate of Ref. \cite{strongmoska}, rescaled down by a factor 0.9. The upper, median and lower curves show the effect of including a signal component calculated for a 100 GeV DM particle which annihilates into a $W^{+}W^{-}$ final state and for a NFW halo profile \cite{nfw} with a scale radius of
20 kpc. A boost factor of 32 is included. The median curve refers to the best--fit set of propagation parameters \cite{parfit}, while the upper and lower curves delimit the uncertainty band on the theoretical determination of the flux. The experimental points refer to the combined 1994, 1995 and 2000 HEAT results \cite{heat}.}\label{fig3} 
\end{figure}

\begin{figure}[t]
\begin{center}
\noindent
\includegraphics [width=0.44\textwidth]{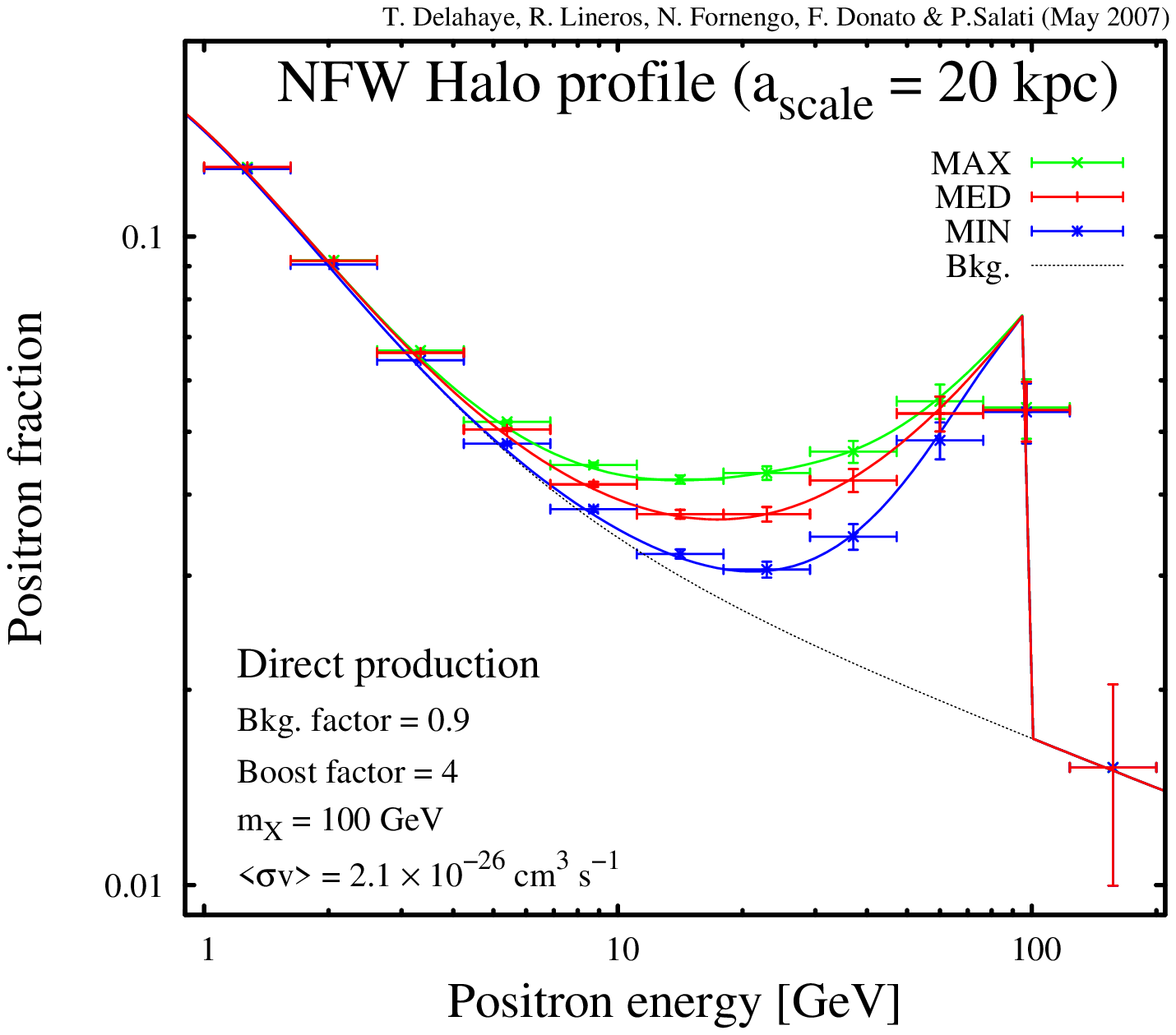}
\end{center}
\vskip -0.75cm
\caption{Predictions for the positron fraction $e^{+}/(e^{+}+e^{-})$ as a function of the positron energy for the PAMELA flight and for a DM particles which annihilates directly into $e^{+}e^{-}$. The lower solid curve refers to background only \cite{strongmoska}. The median curve refers to the best--fit prediction, while the upper and lower curves show the results for the extreme values of the propagation parameters. The error bars denote the predictions for a detector acceptance of 20.5 cm$^{2}$ sr \cite{pamela} and a 3 years flight.}\label{fig4}
\end{figure}

The solution of the diffusion equation allows us now to calculate the positron energy spectra at the Earth, and to determine the uncertainties on the fluxes. One example is given in Fig.\ref{fig2},
where the positron flux for a DM particle which annihilates purely into a $\bar{\tau}\tau$ final state is shown. The annihilation cross section, here and in the following, is chosen to match the relic
abundance for cold DM, as determined by the latest WMAP data \cite{wmap3yr}, under the hypothesis of dominant s--wave annihilation for the DM candidate. The figure shows the band of allowed values for
the positron signal in this channel: the uncertainty ranges from a factor of 5 at energies around 1 GeV down to a factor of 2 above 10 GeV. 

A comparison with the HEAT data \cite{heat}, in terms of the positron fraction, is given in Fig.\ref{fig3}, for annihilation into a $W^{+}W^{-}$ final state by a 100 GeV DM particle.
Due to the contribution of the background \cite{strongmoska} (which we have reduced by 10\%), the uncertainty on the predictions for the total flux is limited to a factor of the order of 20--30\%.
Clearly a complete understanding of the fluctuations of the theoretical predictions on the total positron fraction arising from astrophysical origin will require a study of the background in the same theoretical framework adopted for the signal.
This would allow also to correlate the determination of the two positron components in the comic rays. This is beyond the scope of this paper, which concentrates on the determination of the uncertainty arising specifically from the signal,
and will be addressed elsewhere.
From the same figure we also notice, as shown by other authors,
that the HEAT excess around a positron energy of 10--30 GeV could
be explained by DM annihilations into $W^{+}W^{-}$ pairs,
provided that a boosted DM density is present. In this analysis,
we furthermore reach the robust conclusion that this agreement holds
whatever the cosmic ray propagation model. On the other hand, the HEAT data
alone cannot be used to discriminate between the various possible propagation
models.
However, this discrimination will become possible with the upcoming PAMELA \cite{pamela} experiment.
The predictions for a 3 years flight, in case of a HEAT excess explained by a 100 GeV DM annihilation into a $e^{+}e^{-}$ line, 
are presented in Fig.\ref{fig4}. Should PAMELA detect a feature like the one shown in Fig.\ref{fig4},
this would be a clear signature of a DM signal (a direct production of positrons, in this case) and moreover
the experiment would have potentially the capability to distinguish among different propagation models.
Similar conclusions, with smaller error bars, will be obtained by the AMS experiment \cite{ams} on the International Space Station.

\section{Conclusions}

In this paper we have discussed the positron indirect detection signal of DM annihilation in the galactic halo, focussing our attention to the determination of the astrophysical uncertainties on the positron flux due to the positron propagation inside the galactic medium. We have shown that the uncertainties may be large: for a 100 GeV DM particle, they are of the order of 10--30 for a few GeV positron, and a factor of 2--5 above 10 GeV, depending on the propagation parameters, the halo profile and the injection spectra. The comparison with current data shows that agreement between the predictions and the possible HEAT excess is possible, for DM annihilating mostly into gauge bosons or directly into a positron--electron pair, and the agreement is not limited by the astrophysical uncertanties. We then showed that upcoming experiments, like the PAMELA mission, will have the capabilities, in the case the excess is confirmed and more accurately determined, to shed light also on the astrophysical models underlying galactic propagation.

\section{Acknowledgements}
N.F., R.L. and F.D. gratefully acknowledge research Grants funded jointly by the Italian Ministero dell'Istruzione, dell'Universit\`a e della Ricerca (MIUR),
by the University of Torino and by the Istituto Nazionale di Fisica Nucleare (INFN) within the {\sl Astroparticle Physics Project}.
R.L. also acknowledges the Comisi\'on Nacional de Investigaci\'on Cient\'ifica y Tecnol\'ogica (CONICYT) of Chile.
T.D. acknowledges financial support from the French Ecole Polytechnique and P.S. is
grateful to the French Programme National de Cosmologie.

\bibliography{icrc0186}
\bibliographystyle{plain}

\end{document}